\documentclass[sigconf]{acmart}
\renewcommand\footnotetextcopyrightpermission[1]{} 
\usepackage{listings}
\usepackage{verbatim}
\usepackage{pifont}
\usepackage{subfig}
\usepackage{url} 
\usepackage{fancyvrb}
\usepackage{textcomp}
\usepackage{fancyhdr}
\usepackage{colortbl}

\AtBeginDocument{%
  \providecommand\BibTeX{{%
    \normalfont B\kern-0.5em{\scshape i\kern-0.25em b}\kern-0.8em\TeX}}}

\setcopyright{none}
\copyrightyear{2021}
\acmYear{2021}
\acmDOI{10.1145/1122445.1122456}

\acmConference[Preprint]{}{November}{2021}
\acmBooktitle{}
\acmPrice{15.00} 
\acmISBN{978-1-4503-XXXX-X/18/06}

\begin{document}

\title{Choose Your Programming Copilot}
\subtitle{A Comparison of the Program Synthesis Performance of GitHub Copilot and Genetic Programming}

\author{Dominik Sobania}
\authornote{Both authors contributed equally.}
\affiliation{%
  \institution{Johannes Gutenberg University}
  \city{Mainz}
  \country{Germany}}
\email{dsobania@uni-mainz.de}

\author{Martin Briesch}
\authornotemark[1]
\affiliation{%
  \institution{Johannes Gutenberg University}
  \city{Mainz}
  \country{Germany}}
\email{briesch@uni-mainz.de}

\author{Franz Rothlauf}
\affiliation{%
  \institution{Johannes Gutenberg University}
  \city{Mainz}
  \country{Germany}}
\email{rothlauf@uni-mainz.de}

\renewcommand{\shortauthors}{Sobania, Briesch, and Rothlauf}

\begin{abstract}
GitHub Copilot, an extension for the Visual Studio Code development environment powered by the large-scale language model Codex, makes automatic program synthesis available for software developers. 
This model has been extensively studied in the field of deep learning, however, a comparison to genetic programming, which is also known for its performance in automatic program synthesis, has not yet been carried out.
In this paper, we evaluate GitHub Copilot on standard program synthesis benchmark problems and compare the achieved results with those from the genetic programming literature. In addition, we discuss the performance of both approaches.
We find that the performance of the two approaches on the benchmark problems is quite similar, however, in comparison to GitHub Copilot, the program synthesis approaches based on genetic programming are not yet mature enough to support programmers in practical software development. Genetic programming usually needs a huge amount of expensive hand-labeled training cases and takes too much time to generate solutions. Furthermore, source code generated by genetic programming approaches is often bloated and difficult to understand.
For future work on program synthesis with genetic programming, we suggest researchers to focus on improving the execution time, readability, and usability.
\end{abstract}



\maketitle


\section{Introduction}

In software development, programmers today usually have support from tools such as automatic code completion or comprehensive online resources, 
which greatly accelerate their daily work. Automatic program synthesis, in which source code is generated based on a given definition, e.g., in 
the form of a natural language description or input/output examples \cite{gulwani2010dimensions}, has the potential to become another standard tool in software development.

A tool that has recently drawn some attention is GitHub Copilot\footnote{\url{https://copilot.github.com/}}, an extension for the
Visual Studio Code development environment that offers suggestions for extending a programmer's source code based on problem descriptions and 
existing code. Based on the large-scale language model Codex \citep{chen2021evaluating}, which was trained on a large amount of source code, GitHub Copilot is more than 
a standard code completion tool (which often only suggests variable or function names) as it recommends the source code of complete functions and 
even suggests useful test cases for existing functions. 

Genetic programming (GP) \cite{cramer1985representation, koza1992genetic} is another approach that has made great progress in the field of automatic program synthesis in recent years. 
Starting with a random population of programs, GP applies an evolutionary process to gradually improve the programs and to finally come up with solutions 
that meet the requirements. To define the functionality of the desired programs and to evaluate the generated programs during evolution, 
usually input/output examples are used. 

To make different program synthesis approaches comparable, Helmuth et al. \cite{helmuth2015psb1, helmuth2021psb2} recently curated two benchmark collections containing a wide range of program 
synthesis benchmark problems with different complexity. In addition to a description of the problems, the benchmark suites define also how the
training and test data should be defined. However, the problems of the benchmark suites have so far mainly been used to test and compare different
GP-based program synthesis approaches \cite{sobania2021recent}. A comparison with program synthesis approaches based on large-scale language models has not yet been
carried out.

\begin{figure*}[!ht]
  \centering
  \includegraphics[width=\linewidth]{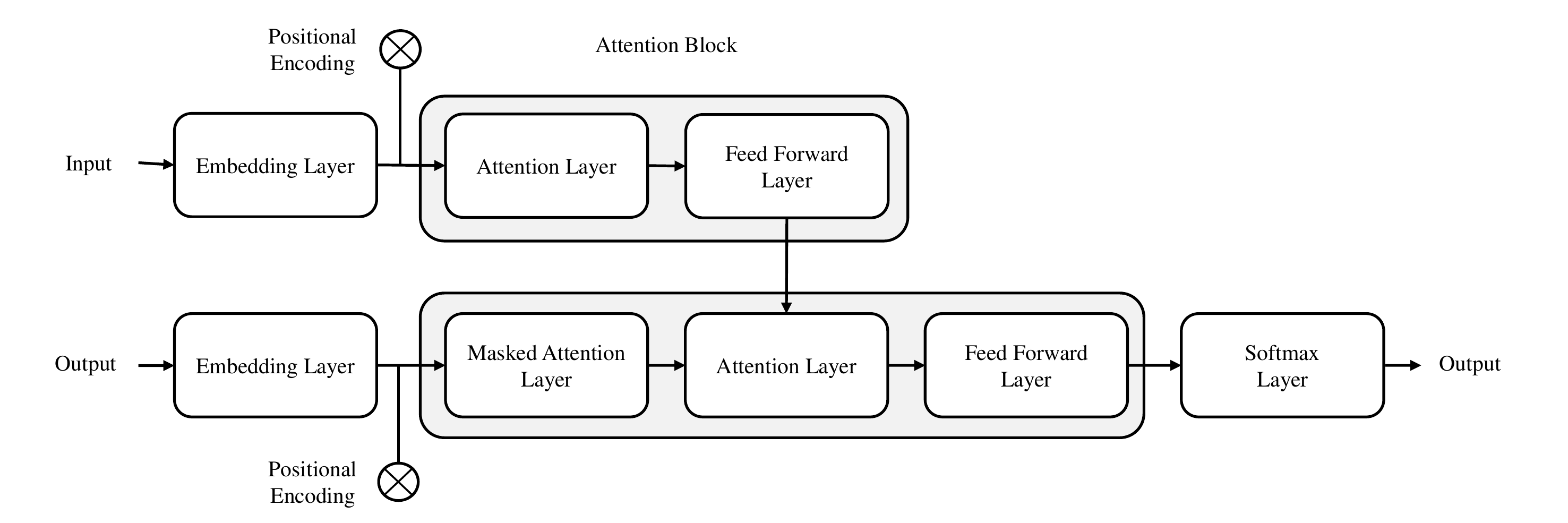}
  \caption{Encoder-Decoder architecture for a typical transformer model. After an embedding layer the input gets fed into multiple attention blocks of the encoder. The output of the encoder, as well as the current output of the whole model, is then processed by multiple attention blocks in the decoder to produce the final output (based on Vaswani et al.~\citep{vaswani2017attention}).}
  \label{fig:transformer}
\end{figure*}

Therefore, in this work, we evaluate GitHub Copilot on the common program synthesis benchmark problems suggested by 
Helmuth et al. \cite{helmuth2015psb1, helmuth2021psb2} and
compare the obtained results with those reported in the GP literature. In addition, we discuss the performance of the GP-based approaches and GitHub
Copilot and identify future research directions. 

To evaluate GitHub Copilot on the benchmark problems, we use the Copilot
Extension\footnote{\url{https://marketplace.visualstudio.com/items?itemName=GitHub.copilot}} 
for Visual Studio Code and provide for every problem the function's signature and the textual problem description as comment. After that, 
we evaluate Copilot's suggestion. If the suggested program is not fully correct, we let Copilot generate further alternatives 
(a maximum of ten) and check these alternative programs for correctness. For GP, we distinguish between the two benchmark sets. For the older program synthesis benchmark suite PSB1, which was published in 2015 \cite{helmuth2015psb1}, we took the GP performance results from a recent survey on program synthesis \cite{sobania2021recent}; for the new benchark suite PSB2, which was published in 2021 \cite{helmuth2021psb2}, we report GP results from all publications dealing with PSB2. 

Following this introduction, we present in Sect.~\ref{sec:related_work} recent work on GP-based program synthesis and briefly introduce 
large-scale language models as used by GitHub Copilot. Section~\ref{sec:experiments} presents the results of our comparison. In 
Sect.~\ref{sec:discussion}, we discuss our findings and identify future research directions. Section~\ref{sec:conclusions} concludes the paper. 

\section{Related Work}\label{sec:related_work}

In this section, we present recent work on GP-based program synthesis and briefly introduce relevant approaches. Furthermore, we describe the foundations of large-scale language models and present work where such
models are used for program synthesis.  

\subsection{GP-Based Program Synthesis}

The publication of PSB1 \cite{helmuth2015psb1} in 2015 revived and consolidated the field of general program synthesis in GP. Different papers became comparable through a common base of benchmark problems, since in addition to the problems, the training and test data (input/output examples) are also defined in the benchmark suite. The benchmark suite PSB2 \cite{helmuth2021psb2} from 2021 extends the collection of common problems.
For solving these benchmark problems, three major categories of methods emerged from the GP field: stack-based GP, grammar-guided GP, and linear GP. 

In order to be able to represent and process programs, a GP approach has to deal with different data types. Stack-based GP approaches provide stacks to differentiate between the supported data types \cite{spector2002genetic}. The stack-based approach mainly used in the literature is PushGP \cite{spector2002genetic, spector2005push3}, which is based on
the stack-based programming language Push. With the introduction of the Uniform Mutation by Addition and Deletion (UMAD) operator \cite{helmuth2018program} and the usage of different selection methods \cite{helmuth2020benchmarking}, the success rates of PushGP in program synthesis could be increased significantly. 

A disadvantage of approaches like PushGP is that stack-based programming languages are not used
in real-world software development. Thus, the generated source code cannot be used directly
in existing software projects. With grammar-guided GP approaches it is possible to 
represent common programming languages like Python \cite{forstenlechner2017grammar, forstenlechner2018extending}. Using a context-free grammar allows, e.g., the usage of different data types, loops, and conditionals. 
To improve grammar-guided GP approaches, Hemberg et al.~\cite{hemberg2019domain} used not only the GP-typical input/output examples, but also the textual problem descriptions of the benchmark problems. Other grammar-guided GP work used source code mined from GitHub to improve the code quality of the programs generated by GP \cite{sobania2019teaching}.  

In linear GP, the provided functions usually operate on data registers similar to those known from common Assembler languages \cite{brameier2007linear}. To reference these registers, Lalejini and Ofria \cite{lalejini2019tag} suggested tag-based memory to make the program representation more stable against changes by variation operators. 

For a broader introduction to the relevant GP approaches for program synthesis and an overview of their success on the common
benchmarks, we refer the reader to a recent literature survey \cite{sobania2021recent}.

\subsection{Large-Scale Language Models}
As an alternative to providing input/output examples to specify the program, it is also possible to formulate the desired behaviour with a natural language description \cite{gulwani2010dimensions}. Natural language processing has seen a variety of improvement in recent years. Much of those improvements can be attributed to the availability of larger data sets, more compute and novel deep learning architectures \citep{vaswani2017attention, radford2018improving, kenton2019bert, NEURIPS2020_1457c0d6}. 

One of the highly influential deep learning  architectures that can be used for natural language processing is the Transformer model \citep{vaswani2017attention}. In contrast to prior architectures like Long Short-Term Memory networks \citep{hochreiter1997long} and Gated Recurrent Unit networks \citep{cho2014learning}, Transformer models do not rely on recurrence and the sequential modelling of recurrent architectures. Instead they solely use attention mechanisms \citep{DBLP:journals/corr/BahdanauCB14} to model dependencies within a sequence. This enables models to be trained more efficiently and on longer sequences while still achieving state-of-the-art results \citep{vaswani2017attention}.

Attention mechanisms allow a model to identify the relevant context by comparing a query of an element at position $i$ in a sequence with every other element of that sequence, also called self-attention. This is done using an attention function which parameters are learned during training.
The most common attention functions are additive attention \citep{DBLP:journals/corr/BahdanauCB14} and dot-product attention \citep{luong2015effective}. 
 
Figure~\ref{fig:transformer} shows the general building blocks of a Transformer model using an Encoder-Decoder architecture. The inputs get encoded with an embedding layer and a positional encoding and are then fed into the attention blocks of the encoder. The encoder consists of multiple stacked attention blocks, each with an attention layer followed by a feed forward layer. The decoder takes the output of the encoder as well as the preceding output of the model as input and also consists of multiple attention blocks. To prevent the model from looking at subsequent input tokens during training, the first attention layer uses masking. Following the last attention block of the decoder the model returns the probabilities for the next token using a softmax layer \citep{vaswani2017attention}.

The attention layers utilize self-attention and a scaled dot-product attention function which is defined as

\begin{equation*}
    {Attention}(Q,K,V) = {softmax}(\frac{QK^T}{\sqrt{d_k}}) V,
\end{equation*}

where $Q$ is the matrix of query vectors $q_i$, $K$ is the matrix of key vectors $k_i$, and $V$ is a matrix of value vectors $v_i$. The vectors $q_i$, $k_i$, and $v_i$ are learned linear projections of the input position $i$ of the previous layer. $\sqrt{d_k}$ is a scaling factor for easier training. Transformers typically use multiple attention functions per layer with different parameters, allowing the network to focus on different aspects of context \citep{vaswani2017attention}.

Since the training of these Transformers can be well parallelized on modern hardware, it is possible to train them efficiently with large data sets.
While large labeled data sets are typically expensive to obtain for specific tasks, it is also possible to pre-train those models in an unsupervised fashion and fine-tune them for a specific task with a much smaller data set \citep{howard2018universal,radford2018improving}. The use of pre-trained models that have a high capacity and are trained with huge data sets, led to a large performance increase across many natural language processing tasks \citep{NEURIPS2020_1457c0d6}.

Given the large amount of freely available open source code, there is much data available to (pre-)train those large-scale language models for program synthesis. Trained models such as CodeBERT \citep{feng2020codebert}, PyMT5 \citep{clement2020pymt5} and Codex \citep{chen2021evaluating} take a description of the desired program behaviour as natural language specification and sample a corresponding Python program. Codex is a model based on the GPT-3 \citep{NEURIPS2020_1457c0d6} architecture with up to 12 billion parameters. This model is first pre-trained with 159 GB of code samples from sources like GitHub in an unsupervised manner. Afterwards, the authors fine-tuned the model with a smaller supervised data set containing functions from competitive programming websites and curated code repositories. The final model achieves promising results \citep{chen2021evaluating}. GitHub Copilot is a special production version of Codex, which we use for our analysis in the following sections.

\section{Performance Comparison}\label{sec:experiments}

In order to be able to compare the quality of GitHub Copilot with the program synthesis performance of GP, we apply GitHub Copilot to the common program synthesis benchmark problems and compare the achieved results to those from the GP-based program synthesis literature. In this section, we describe our test procedure for GitHub Copilot and present our comparison including all benchmark problems from PSB1 and PSB2. 

\definecolor{lightgray}{gray}{0.9}
\begin{table*}
  \caption{
  Performance of GitHub Copilot compared to the results reported in the literature for stack-based GP, grammar-guided GP, and linear GP on the benchmark problems from PSB1. 
  A checkmark (\ding{51}) indicates that the first returned solution is correct (for GitHub Copilot) or there is at least one run reported in the literature returning a successful solution (for GP). A checkmark with an asterisk (\ding{51}*) indicates that not the first solution suggested by GitHub Copilot is correct, but a correct solution can be found in the first ten suggestions. 
  A cross (\ding{55}) indicates that none of the suggested
  programs were successful (Copilot) or the literature reports no found successful solution (GP approaches).}
  \renewcommand{\arraystretch}{1.3}
  \label{tab:psb1_results}
  \begin{tabular}{lcccc}
    \toprule
    \textbf{Benchmark Problem }& \;\;\;\;\;\textbf{GitHub Copilot}\;\;\;\;\; & \;\;\;\;\;\textbf{Stack-based GP}\;\;\;\;\; & 
    \;\;\;\;\;\textbf{Grammar-Guided GP}\;\;\;\;\; & \;\;\;\;\;\textbf{Linear GP}\;\;\;\;\; \\
    \midrule
    Number IO                  & \ding{51}   & \ding{51} & \ding{51} & \ding{51} \\
    \rowcolor{lightgray}
    Small or Large             & \ding{51}   & \ding{51} & \ding{51} & \ding{51} \\
    For Loop Index             & \ding{51}   & \ding{51} & \ding{51} & \ding{51} \\
    \rowcolor{lightgray}
    Compare String Lengths     & \ding{51}   & \ding{51} & \ding{51} & \ding{51} \\
    Double Letters             & \;\ding{51}* & \ding{51} & \ding{51} & \ding{55} \\
    \rowcolor{lightgray}
    Collatz Numbers            & \;\ding{51}* & \ding{55} & \ding{55} & \ding{55} \\
    Replace Space with Newline & \;\ding{51}* & \ding{51} & \ding{51} & \ding{55} \\
    \rowcolor{lightgray}
    String Differences         & \;\ding{51}* & \ding{51} & \ding{55} & \ding{55} \\
    Even Squares               & \ding{51}   & \ding{51} & \ding{51} & \ding{55} \\
    \rowcolor{lightgray}
    Wallis Pi                  & \ding{55}   & \ding{55} & \ding{55} & \ding{55} \\
    String Lengths Backwards   & \ding{51}   & \ding{51} & \ding{51} & \ding{55} \\
    \rowcolor{lightgray}
    Last Index of Zero         & \;\ding{51}* & \ding{51} & \ding{51} & \ding{55} \\
    Vector Average             & \ding{51}   & \ding{51} & \ding{51} & \ding{55} \\
    \rowcolor{lightgray}
    Count Odds                 & \ding{51}   & \ding{51} & \ding{51} & \ding{55} \\
    Mirror Image               & \ding{51}   & \ding{51} & \ding{51} & \ding{55} \\
    \rowcolor{lightgray}
    Super Anagrams             & \ding{55}   & \ding{51} & \ding{51} & \ding{55} \\
    Sum of Squares             & \ding{51}   & \ding{51} & \ding{51} & \ding{55} \\
    \rowcolor{lightgray}
    Vectors Summed             & \ding{51}   & \ding{51} & \ding{51} & \ding{55} \\
    X-Word Lines               & \ding{51}   & \ding{51} & \ding{51} & \ding{55} \\
    \rowcolor{lightgray}
    Pig Latin                  & \ding{55}   & \ding{55} & \ding{51} & \ding{55} \\
    Negative To Zero           & \ding{51}   & \ding{51} & \ding{51} & \ding{55} \\
    \rowcolor{lightgray}
    Scrabble Score             & \;\ding{51}* & \ding{51} & \ding{51} & \ding{55} \\
    Word Stats                 & \ding{55}   & \ding{55} & \ding{55} & \ding{55} \\
    \rowcolor{lightgray}
    Checksum                   & \;\ding{51}* & \ding{51} & \ding{51} & \ding{55} \\
    Digits                     & \ding{55}   & \ding{51} & \ding{51} & \ding{55} \\
    \rowcolor{lightgray}
    Grade                      & \;\ding{51}* & \ding{51} & \ding{51} & \ding{51} \\
    Median                     & \;\ding{51}* & \ding{51} & \ding{51} & \ding{51} \\
    \rowcolor{lightgray}
    Smallest                   & \;\ding{51}* & \ding{51} & \ding{51} & \ding{51} \\
    Syllables                  & \ding{51}    & \ding{51} & \ding{51} & \ding{55} \\
    \midrule
    \pmb{$\Sigma$}\textbf{ (Solved)} & \textbf{24}          & \textbf{25}         & \textbf{25}        & \textbf{7} \\
    \bottomrule
  \end{tabular}
\end{table*}

\subsection{Methodology}

GitHub Copilot is accessible via an extension for the Visual Studio Code development environment. The extension can do both, 
make suggestions for extending existing code (e.g., to complete a single line of code) and suggest complete functions based on a given description. We use the problem descriptions from the PSB1 \cite{helmuth2015psb1} and PSB2 \cite{helmuth2021psb2} papers and let GitHub Copilot suggest 
a function to solve these problems. Therefore, we insert a benchmark problem's textual description as Python comment and define the function's signature adjusted to the considered problem (defining input parameters). Then, we let GitHub Copilot make a suggestion for the function's body (based on the problem's textual description). 

\begin{figure}[!ht]
\begin{lstlisting}[frame=lines,language=Python,numbers=left,xleftmargin=2.4em,framexleftmargin=2.3em,basicstyle=\small]
# Given a string, convert each character in
# the string into its integer ASCII value, 
# sum them, take the sum modulo 64, add the 
# integer value of the space character, and 
# then convert that integer back into its 
# corresponding character (the checksum 
# character). The program must print 
# "Check sum is X", where X is replaced 
# by the correct checksum character.
def myfunc(str1: str):
    # To be completed by GitHub Copilot
\end{lstlisting}
\caption{An example definition as input for GitHub Copilot using the textual description for the Checksum problem. The problem
description was taken from the PSB1 paper \cite{helmuth2015psb1}.}
\label{fig:function_stub}
\end{figure}

Figure~\ref{fig:function_stub} shows an example definition for the Checksum problem. The code lines 1-9 contain the problem description as Python comment and line 10 defines the function's signature. Beginning from line 11, we expect from GitHub Copilot to make a suggestion. 

We assume that GitHub Copilot makes a correct suggestion (is successful) if either the first suggestion or one of the given alternatives 
correctly solves the problem at hand. In our experiments, we allowed a maximum of ten alternatives. 
For the GP-based approaches, we do not carry out any new experiments, but instead use the results from the literature. 

\subsection{Results}

First, we study the results for the benchmark problems suggested in PSB1. We compare the results obtained by GitHub Copilot in our experiments with the results obtained by stack-based GP, grammar-guided GP, as well as linear GP approaches from the literature. The results for the GP approaches are taken from \cite{sobania2021recent}. This recent literature review found $54$ papers which study program synthesis using problems from PSB1. Out of these, $30$ papers reported success on some benchmark problems for stack-based GP, $11$ for grammar-guided GP, and $4$ for linear GP. A test problem is solved by a GP approach, if there is at least one paper that finds a successful solution for the problem at hand in at least one run (independently of the overall number of runs). 

\begin{figure}[!ht]
\begin{lstlisting}[frame=lines,language=Python,numbers=left,xleftmargin=2.4em,framexleftmargin=2.3em,basicstyle=\small] 
def myfunc(str1: str):
    sum = 0
    for i in range(len(str1)):
        sum += ord(str1[i])
    sum = sum % 64
    sum += ord(' ')
    return chr(sum)
\end{lstlisting}
\caption{First suggestion of GitHub Copilot in our experiments for the Checksum problem. The result is not printed but returned. }
\label{fig:checksum_return}
\end{figure}

\begin{figure}[!ht]
\begin{lstlisting}[frame=lines,language=Python,numbers=left,xleftmargin=2.4em,framexleftmargin=2.3em,basicstyle=\small]
def myfunc(str1: str):
    sum = 0
    for char in str1:
        sum += ord(char)
    checksum = chr(sum % 64 + ord(" "))
    print("Check sum is {}".format(checksum))
\end{lstlisting}
\caption{One of the alternatives suggested by GitHub Copilot for the Checksum problem. The result is printed in the correct format. }
\label{fig:checksum_print}
\end{figure}

Table~\ref{tab:psb1_results} shows the results obtained by GitHub Copilot compared to the results reported in the literature for stack-based GP, grammar-guided GP, and linear GP on the benchmark problems from PSB1. A checkmark (\ding{51}) indicates that a successful solution has been found with the first suggestion (for GitHub Copilot) or has been reported in the literature (for the GP approaches). A checkmark with an asterisk (\ding{51}*) indicates that a correct solution was not found with GitHub Copilot's first suggestion but in one of the suggested alternatives (a maximum of ten). 
By making this distinction, we take into account the standard use of GitHub Copilot (and other code completion tools) by a programmer, because a programmer would usually prefer using the first suggestion before studying the alternatives.
A cross (\ding{55}) indicates that none of the suggested programs were successful (Copilot) or the literature reports no found successful solution (GP approaches).

\begin{table}
  \caption{Performance of GitHub Copilot compared to GP for the benchmark problems from PSB2. A checkmark (\ding{51}) indicates that the first returned solution is correct (for GitHub Copilot) or there is at least one GP approach in the literature where at least one GP run returns a successful solution (for GP). A checkmark with an asterisk (\ding{51}*) indicates that not the first solution suggested by GitHub Copilot is correct, but a correct solution can be found in the first ten suggestions. A cross (\ding{55}) indicates that none of the ten suggested
  programs were successful (for Copilot) or the literature reports no found successful solution (for GP).}
  \renewcommand{\arraystretch}{1.3}
  \label{tab:psb2_results}
  \begin{tabular}{lcc}
    \toprule
    \textbf{Benchmark problem}\;\; & \;\;\;\;\textbf{GitHub Copilot}\;\;\;\; & \;\;\;\;\textbf{GP}\;\;\;\; \\
    \midrule
    Basement                    & \ding{51}     &  \ding{51} \\
    \rowcolor{lightgray}
    Bouncing Balls              & \ding{55}     &  \ding{51} \\
    Bowling                     & \ding{55}     &  \ding{55} \\
    \rowcolor{lightgray}
    Camel Case                  & \ding{55}     &  \ding{51} \\
    Coin Sums                   & \ding{51}     &  \ding{51} \\
    \rowcolor{lightgray}
    Cut Vector                  & \ding{55}     &  \ding{55} \\ 
    Dice Game                   & \ding{55}     &  \ding{51} \\
    \rowcolor{lightgray}
    Find Pair                   & \ding{51}     &  \ding{51} \\
    Fizz Buzz                   & \ding{51}     &  \ding{51} \\
    \rowcolor{lightgray}
    Fuel Cost                   & \ding{51}     &  \ding{51} \\
    GCD                         & \ding{51}     &  \ding{51} \\
    \rowcolor{lightgray}
    Indices of Substring        & \ding{55}     &  \ding{51} \\
    Leaders                     & \ding{55}     &  \ding{55} \\
    \rowcolor{lightgray}
    Luhn                        & \;\ding{51}*   &  \ding{55} \\
    Mastermind                  & \ding{55}     &  \ding{55} \\
    \rowcolor{lightgray}
    Middle Character            & \ding{51}     &  \ding{51} \\
    Paired Digits               & \ding{51}     &  \ding{51} \\
    \rowcolor{lightgray}
    Shopping List               & \ding{51}     &  \ding{55} \\
    Snow Day                    & \ding{55}     &  \ding{51} \\
    \rowcolor{lightgray}
    Solve Boolean               & \ding{55}     &  \ding{51} \\
    Spin Words                  & \ding{51}     &  \ding{55} \\
    \rowcolor{lightgray}
    Square Digits               & \;\ding{51}*   &  \ding{51} \\
    Substitute Cipher           & \ding{51}     &  \ding{51} \\
    \rowcolor{lightgray}
    Twitter                     & \ding{51}     &  \ding{51} \\
    Vector Distance             & \;\ding{51}*   &  \ding{55} \\
    \midrule
    \pmb{$\Sigma$}\textbf{ (Solved)}   & \textbf{15}    & \textbf{17} \\
  \bottomrule
\end{tabular}
\end{table}

Overall, the results for GitHub Copilot and the GP approaches are quite similar. GitHub Copilot finds solutions for $24$ benchmark problems and the GP approaches found up to $25$ (stack-based and grammar-guided GP). For linear GP, the literature reports successful solutions for only $7$ benchmark problems. However, this does not necessarily mean that linear GP generally performs worse than the stack-based and grammar-guided approaches, as there are just a lower number of papers studying linear GP for these benchmark problems, which means that the coverage of the problems is not as high as for stack-based and grammar-guided GP. So it is not surprising that the number of successes is lower for the linear GP approaches. 

From the 29 problems from PSB1, GitHub Copilot does not find successful solutions for $5$ problems: Wallis Pi, Super Anagrams, Pig Latin, Word Stats, and Digits. However, benchmark problems like Wallis Pi and Word Stats seem to be very complex as also there is not one GP approach in the literature that can solve these problems. The reader should be aware that in standard GP papers usually 100 runs are performed and the problem is counted as solved if at least one of the hundred runs returns a successful solution  (i.a., in \cite{helmuth2015psb1}, \cite{forstenlechner2017grammar}, and \cite{sobania2020challenges}). This gives GP approaches a higher chance to find successful solutions. 

For some benchmark problems, GitHub Copilot does not find a fully correct solution with the first suggestion but one of the alternative
solutions is correct (marked by \ding{51}*). However, often the solution suggested first by Copilot is very close to a correct solution. 
For example, PSB1 requires for some benchmark problems that the result should be printed and not simply returned \cite{helmuth2015psb1}. 

Figure~\ref{fig:checksum_return} shows such an example suggestion for the Checksum problem where the result is returned and not printed as
required. Nevertheless, for a programmer this is still a helpful solution as the \texttt{return} statement (line $7$) can be easily replaced with
a \texttt{print()} function. Furthermore, in the suggested alternatives there is often a suggestion that completely fulfills the requirements. 
An example is Figure~\ref{fig:checksum_print}, which shows one of the suggestions of GitHub Copilot for the Checksum problem where the result is printed in the correct format. 

In addition to the benchmark problems from PBS1, we study also the performance of GitHub Copilot on the problems from PSB2. Again, we
compare the obtained results to the results available in the GP literature. This means, we took the results from the two available publications, namely the PSB2 paper \cite{helmuth2021psb2} as well as Helmuth and Spector \citep{helmuth2021problem}.
Since there are so far only two papers that use the problems from PSB2, we do not differentiate this time between the GP method used. 

Table~\ref{tab:psb2_results} shows the results obtained by GitHub Copilot in our experiments compared to the results reported 
in the GP literature on the benchmark problems from PSB2. To mark the results, we use the same notation as in Table~\ref{tab:psb1_results}.

Again, the overall results are quite similar for GitHub Copilot and GP. GitHub Copilot finds a working solution for $15$ and GP for $17$ benchmark problems. The benchmark problems for which a successful solution is found are distributed similarly for GitHub Copilot and GP.
Only for Luhn, Shopping List, Spin Words, and the Vector Distance problem, GitHub Copilot finds a working solution where GP
fails. Conversely, GP finds solutions for Bouncing Balls, Camel Case, Dice Game, Indices of Substring, Snow Day, and the Solve Boolean problem where none of the analyzed suggestions of
GitHub Copilot are fully correct. 
However, on those problems GP only finds a solution in less than 10\% of the runs.

\section{Discussion}\label{sec:discussion}

To better understand our findings, we exemplarily analyze and extend some of the solutions suggested by GitHub Copilot and discuss the implications for GP-based program synthesis. 

\subsection{Analyzing Suggested Solutions}

Generalization is a general challenge for machine learning approaches. For a real-world usage, machine learning methods must also produce high quality results for previously unseen inputs; simply memorizing the training data is not sufficient. Low generalization is a relevant issue in large-scale language models \citep{carlini2021extracting}, and also GP-based approaches suffer from poor generalization for some of the program synthesis benchmark problems \cite{helmuth2017improving, sobania2021generalization_problems}. 

As the problems from the program synthesis benchmark suites are common exercises for programming beginners as well as coding challenges \cite{helmuth2015psb1, helmuth2021psb2}, it can be expected that solutions to some of these problems, albeit often in slightly different versions, can be found in open source repositories and that GitHub Copilot has seen this data during the training process. 

During our experiments we noticed for some benchmark problems that GitHub Copilot has some knowledge about some of the problems. E.g., for the Coin Sums problem, when just entering the first words of the problem description ("Given a number of cents"), GitHub Copilot suggests a continuation of the description as well as a function that has some similarities to the benchmark problem 
from PSB2. Figure~\ref{fig:coin_sum_1} shows the completed problem description and the suggested function.\footnote{For all code examples in this section, we added additional line breaks for better readability.} The suggested function is not a solution for the problem from PSB2, but it illustrates that GitHub Copilot not only has some knowledge about US-Dollar coins (see line 4, Fig.~\ref{fig:coin_sum_1}), but also a bias in the suggestions as the description does not mention that US-Dollar coins are wanted (e.g., Euro coins could be also possible). 
When giving GitHub Copilot the complete problem description from the PSB2 paper (which is more specific and also defines the coins), GitHub Copilot suggests a correct solution which is shown in Fig.~\ref{fig:coin_sum_2}. However, this benchmark problem could also have been included in GitHub Copilot's training data in the exact same form. Thus, can only solutions be suggested for problems that are already known?

\begin{figure}[!t]
\begin{lstlisting}[frame=lines,language=Python,numbers=left,xleftmargin=2.4em,framexleftmargin=2.3em,basicstyle=\small]
# Given a number of cents, return the minimum 
# number of coins needed to make change.
def func(cents):
    coins = [25, 10, 5, 1]
    num_coins = 0
    for coin in coins:
        num_coins += cents // coin
        cents %= coin
    return num_coins
\end{lstlisting}
\caption{An example code where we entered only the first five words of the description of the Coin Sums problem and GitHub Copilot already suggested the continuation of the problem description and the associated function.}
\label{fig:coin_sum_1}
\end{figure}

\begin{figure}[!t]
\begin{lstlisting}[frame=lines,language=Python,numbers=left,xleftmargin=2.4em,framexleftmargin=2.3em,basicstyle=\small]
# Given a number of cents, find the fewest 
# needed number of US coins (pennies, nickels, 
# dimes, and quarters) to make that amount, 
# and return the number of each type of count 
# as a separate output.
def myfunc(int1: int):
    return (int1 // 25, int1 % 25 // 10,
        int1 % 25 % 10 // 5,
        int1 % 25 % 10 % 5)
\end{lstlisting}
\caption{Solution suggested by GitHub Copilot for the Coin Sums problem with the complete textual description from PSB2 \cite{helmuth2021psb2}.}
\label{fig:coin_sum_2}
\end{figure}

To examine if it is possible to create novel functionality with GitHub Copilot, we tested several inputs ranging from descriptions of
arithmetic problems to specific graphical plots. In most cases, GitHub Copilot was able to make useful suggestions. Even 
if the suggestions were not completely correct, they often contain the relevant libraries and function calls a programmer would otherwise look up in online resources. 

Exemplarily, Figure~\ref{fig:checksum_extended} shows GitHub Copilot's suggestion for the Checksum problem with a changed problem description (original problem description from PSB1 \cite{helmuth2015psb1} given in Fig.~\ref{fig:function_stub}). All of our
changes are correctly implemented in Copilot's suggested function: the correct value is used for the modulo of the sum (line 16),
the ASCII values are calculated for the correct letters (lines 17-18), and the intermediary result is multiplied with the correct value (line 19). So GitHub Copilot recognizes relationships between the problem description and the corresponding code lines, even for problems previously unknown. It is also noticeable, that the source code generated by GitHub Copilot could be significantly shorter (e.g., by using an existing \texttt{sum()} function and doing several steps in a single line), but it is still easily readable and understandable. This is often different in GP-based program synthesis as GP may produce uncommon solution strategies and bloated code \cite{sobania2019teaching, sobania2020challenges}. 
 
For benchmark problems from PSB2, we noticed that often the textual problem descriptions are not specific enough to explicitly solve the problem (even for human programmers). This is not an issue for GP-based approaches as they are usually based on the given input/output examples but GitHub Copilot requires an adequate problem description. However, with an adjusted description, a solution can often be found with Copilot. Fig.~\ref{fig:boolean_different} shows the suggested solution for the Solve Boolean
problem if we extend the existing PSB2 problem description (we added an explanation for every operator and terminal). While GitHub Copilot was not able to find a correct solution in our experiments with the original description (see Table~\ref{tab:psb2_results}), a correct solution can be found using a more specific problem description.

Again, the generated source code is easy to read and understand. However, the code snippet reveals an issue of language models 
trained on publicly available source code which may contain insecure source code \citep{rokon2020malware,chen2021evaluating}. 
The example function uses an \texttt{eval()} function (line 10, Fig.~\ref{fig:boolean_different}) without checking the given input string. Such a function in a productive system could pose potential security hazards. In GP-based program synthesis, this is not
an issue as the choices for GP are limited/controlled by a grammar or a function and terminal set. So functions like \texttt{eval()} or \texttt{exec()} can be excluded easily.

\begin{figure}[!t]
\begin{lstlisting}[frame=lines,language=Python,numbers=left,xleftmargin=2.4em,framexleftmargin=2.3em,basicstyle=\small]
# Given a string, convert each character in
# the string into its integer ASCII value, 
# sum them, take the sum modulo 55, add the 
# integer value of the "A" character, then 
# add the integer value of the "Z" character, 
# multiply it by 5, and then convert that 
# integer back into its corresponding 
# character (the checksum character). The 
# program must print "Check sum is X", where
# X is replaced by the correct checksum 
# character.
def myfunc(str1: str):
    sum = 0
    for i in str1:
        sum += ord(i)
    sum = sum % 55
    sum += ord("A")
    sum += ord("Z")
    sum *= 5
    sum = chr(sum)
    print("Check sum is", sum)
\end{lstlisting}
\caption{Solution suggested by GitHub Copilot for the Checksum problem with changed problem description from PSB1 \cite{helmuth2015psb1} (line breaks adjusted).}
\label{fig:checksum_extended}
\end{figure}

\begin{figure}
\begin{lstlisting}[frame=lines,language=Python,numbers=left,xleftmargin=2.4em,framexleftmargin=2.3em,basicstyle=\small]
# Given a string where "T" means True, "F" 
# means false, "|" is the or operator, and "&" 
# is the and operator, return the evaluation 
# of the given string as Boolean value.
def myfunc(str1: str):
    str1 = str1.replace("&", " and ")
    str1 = str1.replace("|", " or ")
    str1 = str1.replace("T", "True")
    str1 = str1.replace("F", "False")
    return eval(str1)
\end{lstlisting}
\caption{Suggested solution for an alternative problem description with additional information for the Solve Boolean problem.}
\label{fig:boolean_different}
\end{figure}

\subsection{Different Approaches, Different Strengths}

In our comparison, GitHub Copilot and the GP-based approaches achieved similar results on the benchmark problems. However, the approaches strongly differ from a user's perspective in the specification of the user's intent. GitHub Copilot usually gets a textual problem description while a GP run uses input/output examples. In a real world use case, it is easier for programmers to define their intent with a textual description than with a large number of input/output examples, as it is expensive to manually generate a set consisting of 100 or more input/output examples \cite{sobania2021generalizability}. For small problems it would be faster to program the function than to create the examples manually. 

On the other hand, when using input/output examples, edge cases can be more easily addressed (like with unit tests 
in real-world software development). However, if there are many input/output examples available and it is difficult to define the problem with a short description text, then GP is currently the better alternative. 
This can also be seen in our results, since GP can also solve some benchmark problems which GitHub Copilot cannot solve as the problem description is not sufficient. However, we should keep in mind, that the shown results for GP are an aggregation of the results reported in a variety of different papers. In practice, a programmer would have to choose one of these approaches, which then may have a lower performance on the benchmark problems. 

Additionally, the response time of a program synthesis approach is important for practical use. Even if the initial training of a large-scale language model is very time consuming and computationally expensive \cite{chen2021evaluating}, GitHub Copilot can suggest solutions in just a few seconds. On the other hand, in GP-based program synthesis, the training process is less expensive, but whenever a new sample of input/output cases is given, the training process has to be repeated. As the current GP-based program synthesis frameworks still may take days to synthesize a solution \cite{helmuth2021psb2}, GP is not yet ready for a real-world usage. However, there is still room for improvement, as hardware-based acceleration (e.g., with GPUs) could significantly reduce the execution time of GP-based program synthesis as shown, e.g., for symbolic regression with GP \cite{baeta2021tensorgp}.

\section{Conclusions}\label{sec:conclusions}

The automatic code completion tool GitHub Copilot has recently drawn some attention for its program synthesis performance. However, 
a comparison with GP, which is also known for its success in automatic program synthesis, has not yet been carried out.

Consequently, we evaluated in this work GitHub Copilot on common program synthesis benchmark problems \cite{helmuth2015psb1, helmuth2021psb2} and compared the obtained results with those from the GP literature. Furthermore, we discussed the performance of the GP-based approaches and GitHub Copilot.

We found that GitHub Copilot and GP perform similar on the studied benchmark problems. Overall, GP can solve more problems, but this comes at the price of practical usage, as GP usually needs many expensive hand-labeled training cases and takes too much time to generate a solution. Furthermore, the suggestions of GitHub Copilot are usually human readable while source code generated by GP is often bloated and difficult to understand. However, GitHub Copilot must be viewed as a black-box, as the user has no information about the exact training data used to generate the model. The generated code could potentially be malicious, biased, or insecure. On the other hand,
with GP it is easily possible to control the training data as well as the choices GP can make during evolution (by designing appropriate grammars or function sets). 

For future program synthesis research with GP, we suggest researchers to focus on improving the execution time, the readability of the generated code, and reducing the amount of required training data. Furthermore, GP must be integrated into the standard software development infrastructure (like GitHub Copilot) in order to be accessible for end-users.

\begin{acks}
The authors would like to thank the GitHub Copilot extension for Visual Studio Code for the helpful discussions on our experimental source code. 
\end{acks}

\bibliographystyle{ACM-Reference-Format}
\bibliography{sample-base} 

\end{document}